\documentclass[%
 reprint,
superscriptaddress,
 amsmath,amssymb,
 aps,
floatfix,
]{revtex4-1}

\usepackage{graphicx}
\usepackage{dcolumn}
\usepackage{bm}

\usepackage{epsfig, graphicx}
\usepackage{subfig}
\usepackage{graphicx}
\usepackage{amsmath}
\usepackage{amssymb}
\usepackage{url}
\usepackage{pgfplots}
\usepackage[percent]{overpic}
\usepackage{caption}
\usepackage{algorithm}
\usepackage{algpseudocode}
\usepackage{natbib}
\begin{document}

\preprint{APS/123-QED}

\title{Clean up or mess up: the effect of sampling biases on measurements of degree distributions in mobile phone datasets}

\author{Adeline Decuyper}
 \altaffiliation[Also at ]{Center for Operations Research and Econometrics, \\ Universit\'e catholique de Louvain, Louvain-La-Neuve, Belgium}
  \email{adeline.decuyper@uclouvain.be}
\author{Arnaud Browet}
\affiliation{
 Institute of Information and Communication Technologies, Electronics and Applied Mathematics\\ Universit\'e catholique de Louvain, Louvain-La-Neuve, Belgium
}

\author{Vincent Traag}

\affiliation{
 CWTS, Leiden University\\
 Leiden, The Netherlands
}
\author{Vincent D. Blondel}
\affiliation{
 Institute of Information and Communication Technologies, Electronics and Applied Mathematics\\ Universit\'e catholique de Louvain, Louvain-La-Neuve, Belgium
}
\author{Jean-Charles Delvenne}
 \altaffiliation[Also at ]{Center for Operations Research and Econometrics, \\ Universit\'e catholique de Louvain, Louvain-La-Neuve, Belgium}
\affiliation{
 Institute of Information and Communication Technologies, Electronics and Applied Mathematics\\ Universit\'e catholique de Louvain, Louvain-La-Neuve, Belgium
}

\date{\today}

\begin{abstract}
Mobile phone data have been extensively used in the recent years to study social behavior. However, most of these studies are based on only partial data whose coverage is limited both in space and time. In this paper, we point to an observation that the bias due to the limited coverage in time may have an important influence on the results of the analyses performed. In particular, we observe significant differences, both qualitatively and quantitatively, in the degree distribution of the network, depending on the way the dataset is pre-processed and we present a possible explanation for the emergence of Double Pareto LogNormal (DPLN) degree distributions in temporal data. 

\end{abstract}

\maketitle


\section{\label{sec:level1}Introduction}

The analysis of social networks has been, for a few years now, a very attractive topic that many researchers have taken interest in. As one particular type of social network, mobile phone graphs have been widely studied in the past decade \citep{blondel2015survey} and researchers have explored the potential to use such data as, among others, sensors to uncover trends of human behavior \citep{karsai2011small, kovanen2013temporal}, mobility habits \citep{gonzalez2008understanding, csaji2013exploring} or social interactions \citep{blondel2010regions, ratti2010redrawing}. However, those studies are inherently based on partial data, often covering a subset of the population, a specific time frame or a single country, as mostly provided by a single telecommunication company. Moreover, the covered subset of the population may be biased, as some providers are more popular across, for example, a given age group, a given scale of revenue, or users preferring voice calls to text messages, to name only a few of the potential sources of bias. Mobile telecommunication companies often do not have a monopolistic position, due to market regulations by the authorities, and the data and the analyses on such data are therefore subject to inherent bias, due to the partial (yet significant) coverage of the population by the telecom operator in the country of interest. \\ 

In particular, the network of mobile phone users of a given operator is subject to changes over time: users can join or leave the network due to various reasons, for example by subscribing a contract with a different operator. The datasets studied in research always cover a given period of time, during which some users appeared and others disappeared from the network. Thus, instead of analyzing the behavior of a fixed group of users during the given time period, studies observe only a partial view of the network and some users are only observed during part of the observation time window. Furthermore, even though we know that this bias exists, it is difficult to remove it. Indeed, it is often very difficult to distinguish between a user that is simply inactive for a few days and a user that has permanently left the network. \\ 

Most of the literature on mobile phone datasets analysis, however, is based on the assumption that the network of users is not biased. Only few studies show the potential bias of their data, and very little is known about the qualitative and quantitative effects that a biased sampling of the users of a country could have on the results of the analyses of a mobile phone network. Moreover, besides this inherent bias of the dataset available, many researchers start by preprocessing the data, often called ``cleaning", removing links or nodes that are not active enough. For example, Onnela \textit{et al.} remove all links that are non-reciprocated, that~is, if an individual $i$ called another user $j$, but $j$ never called $i$, then the link $(i,j)$ is removed from the network \citep{onnela2007structure}. Going even further, Lambiotte \textit{et al.} impose that a pair of nodes $(i,j)$ communicated at least 6 times in each direction for the link to be taken into account in the analyses \citep{lambiotte2008geographical}. These apparently innocuous filtering methods applied before the analysis of the network generate additional biases that are often overlooked. 
In this paper, we show that if we only take into account the users that remain in the network during the whole observation period, thus removing all users that joined or left the network during that time frame, the degree distribution changes from a DPLN to a LogNormal, both in empirical and theoretical scenarios. The limits of validity of this observation are not quite understood and we formulate assumptions to that effect. In that regard, we point to an observation that, far from settling the case of cleaning-induced biases, opens a line of research.

\section{Degree distributions in a mobile phone network}
One of the first things that come to mind when studying social networks is to count the number of acquaintances of each user in the network, referred to as the user's degree. Different studies have shown that the size and shape of the distribution of degrees in a mobile phone network can vary depending on many parameters. In one of the first studies on mobile phone data, looking at one day of data, Aiello \textit{et al.}\ \citep{aiello2000random} observed that the network exhibited a power-law (or Pareto) distribution, corresponding to a random graph model with probability distribution: $ P(\textrm{degree}=x)=\alpha x^{- \beta}$. This observation was later confirmed by studies on different mobile phone datasets \citep{lambiotte2008geographical, onnela2007structure}. However, in a later study by Seshadri \textit{et al.}\ \citep{seshadri2008mobile}, this time using a longer period of one month of data, the authors observed that the mobile call graph had a degree distribution corresponding to a Double Pareto LogNormal (DPLN) \citep{reed2004double}. DPLN distributions are composed of two Pareto distributions, one for small values and one characterizing the tail, joined by a smooth transition and can be derived as a mixture of LogNormal distributions. \\ 

In this paper, we analyze two large databases of mobile phone communications. \\ 
\textbf{Data for Belgium} are recorded over a 6-month period from October 1, 2006 to March 31, 2007 by one large provider in Belgium whose market share is around 30\%. The database contains the communications (SMS and voice calls) of about 3.3M users geographically spread over the whole country. For each call, the information contained in the Call Detail Records (CDRs) is the caller and callee anonymized ID's, the date and time of the communication, whether it is a voice call or SMS, and the duration in case of a voice call. This dataset has already been used in several research projects addressing different questions \citep{krings2012effects, krings2009scaling, krings2009urban, blondel2008fuc, lambiotte2008geographical, blondel2010regions}. \\ 
\textbf{Data for Portugal} are recorded over a 15-month period from April 1, 2006 to June 30, 2007, but with a gap where data is missing between September 16, and October 31, 2006. The data contain all voice calls between clients of the same provider, but SMS are not recorded. The dataset contains information over about 1.9M users, which represent approximately 20\% of the population of the country. For each call, the information recorded is the caller and callee anonymized ID's, the date and time of the communication, and the ID of the cell tower that recorded the call. This dataset has been used before to study mobility patterns in Portugal \citep{csaji2013exploring}. \\ 

We study the distribution of degrees in the network of mobile phone communications. We draw a link between two users as soon as they have communicated at least once, thus without applying any filter on the links. We observe the degree distribution of two networks, namely the network of all users present in the database (hereafter referred to as $G_1$), and another network where only users that are active already from the beginning of the observation period \textit{and} are still active at the end, are taken into account (hereafter referred to as $G_2$). To this end, in this second network $G_2$, we only consider users that are active at least once in the first four weeks, and once in the last four weeks of the observed time period. In the Belgian dataset, these nodes of $G_2$ represent about 70\% of the nodes of $G_1$. In the Portuguese dataset, the nodes that remain active throughout the observation period represent 45\% of the nodes of the whole network. The detailed numbers of nodes are given in Table \ref{tab:ts_numnodes}, along with the numbers of links after 26 weeks of observation, and after 50 weeks in the case of Portugal. \\

\begin{table}[!t]
\caption{number of nodes and number of links in the studied networks.}
\label{tab:ts_numnodes}
\begin{center}
\resizebox{8.6cm}{!}{
\begin{tabular}{lccc}
\hline \hline
 & \#nodes & \#links  & \#links  \\
  & & after 26 weeks & after 50 weeks \\
 \hline
Belgium $ G_1$ & 3,309,113 & 33,610,070  & \\
Belgium $ G_2$ & 2,424,998 & 26,149,120  & \\
\hline
Portugal $G_1$ & 1,944,004 & 12,106,326 & 17,770,357 \\
Portugal $G_2$ & 875,164 & 6,134,970 & 8,565,582 \\
\hline \hline
\end{tabular} }
\end{center}
\end{table}

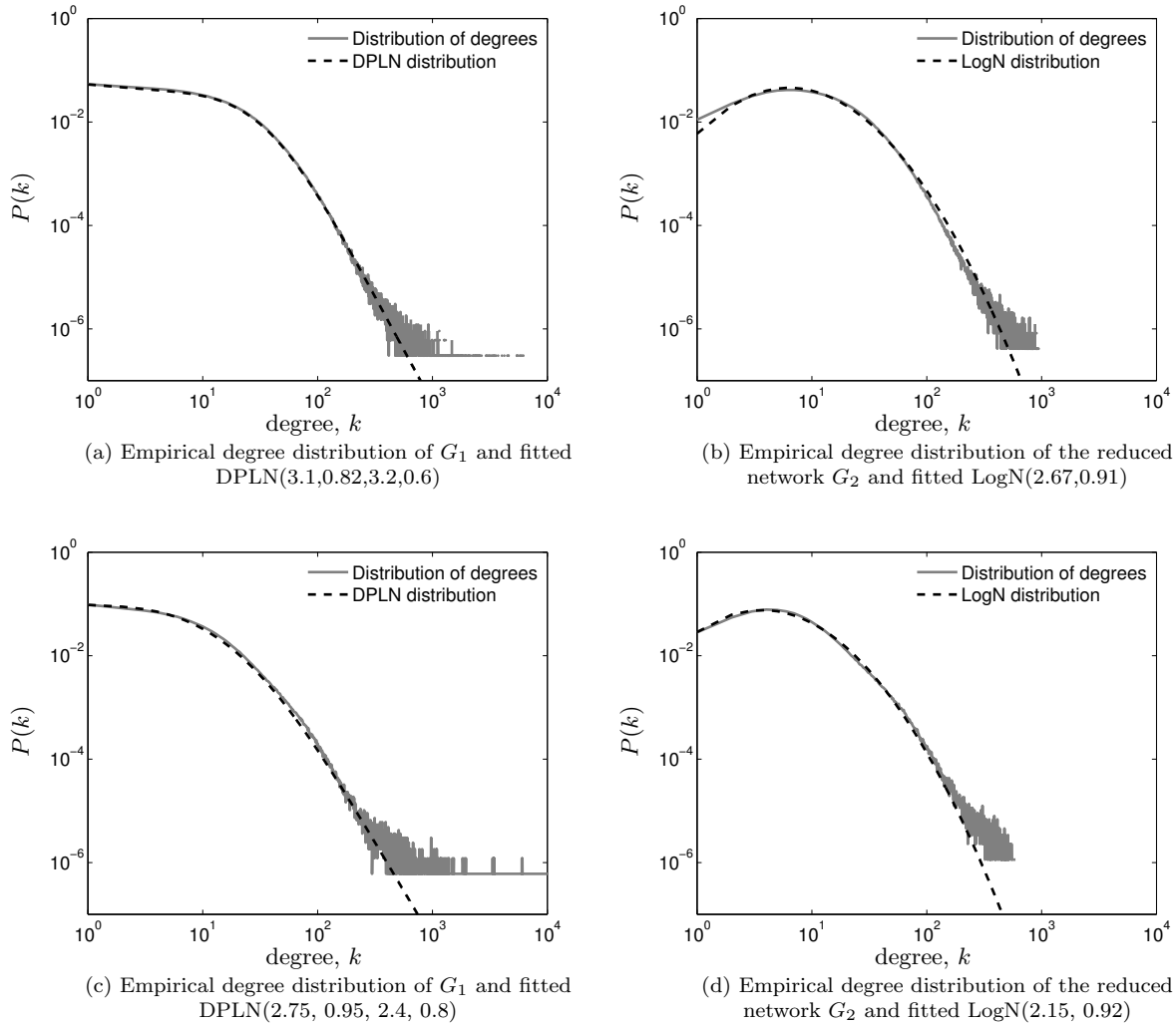
\begin{figure*}[!t]
\centering
\begingroup
\captionsetup[subfigure]{width=7cm}
\subfloat[Empirical degree distribution of $G_1$ and fitted DPLN(3.1,0.82,3.2,0.6)]{
\begin{overpic}[scale=0.5]{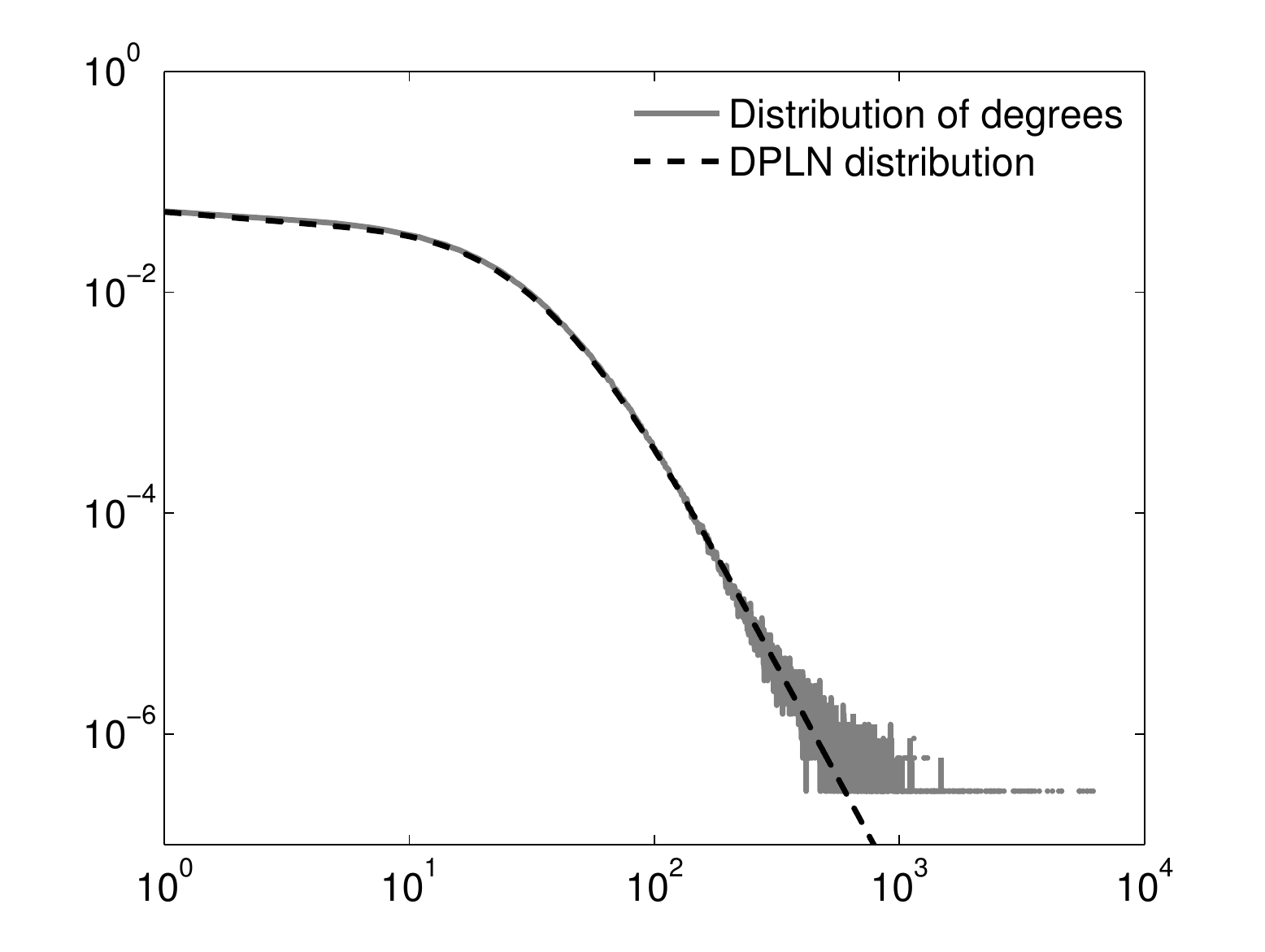}
 \put (43,0) {degree, $k$}
  \put (0,35) {\rotatebox{90}{$P(k)$}}
\end{overpic}
\label{fig:ts_degdplnfit}} 
\subfloat[Empirical degree distribution of the reduced network $G_2$ and fitted LogN(2.67,0.91)]{
\begin{overpic}[scale=0.5]{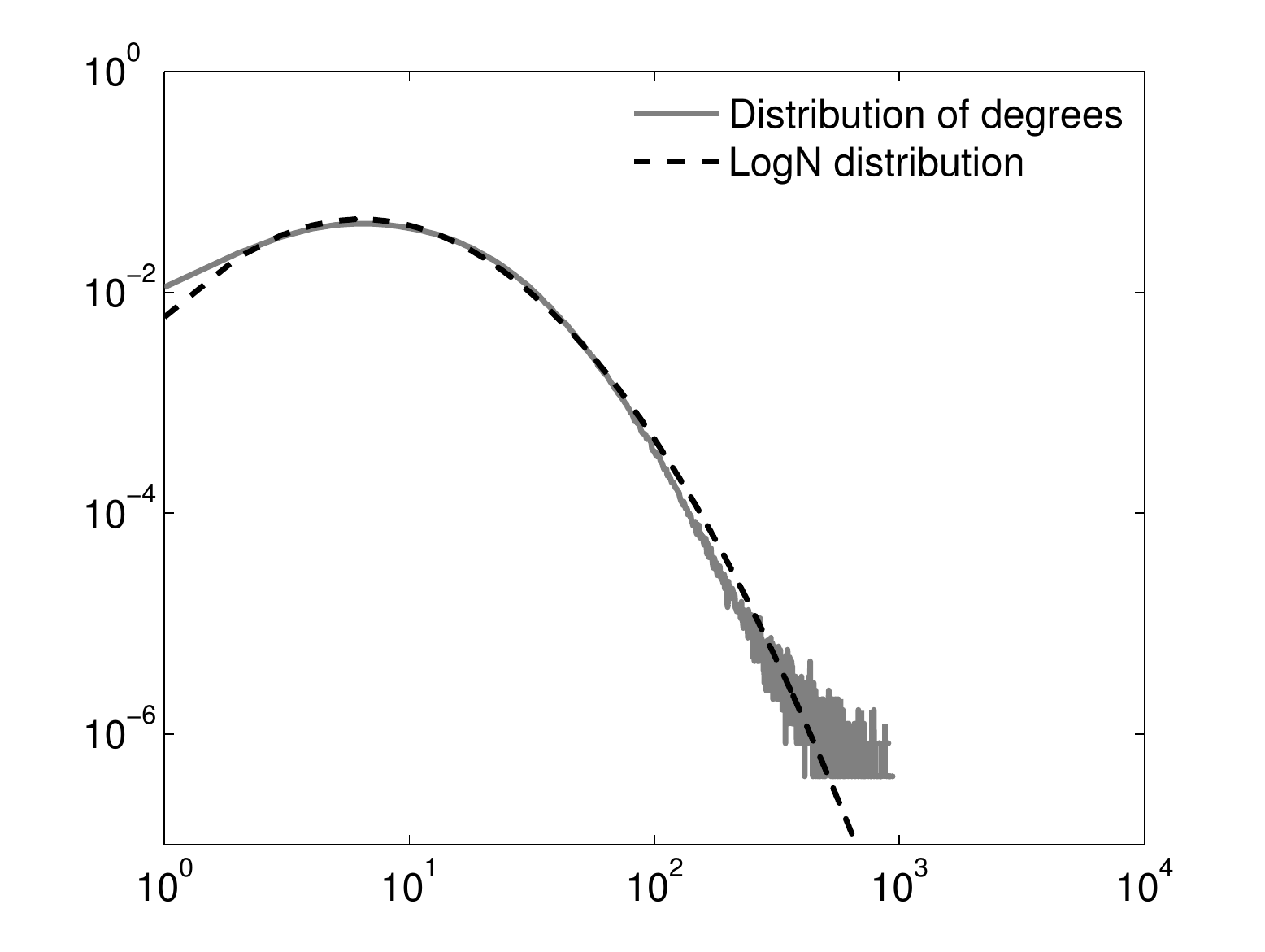}
 \put (43,0) {degree, $k$}
  \put (0,35) {\rotatebox{90}{$P(k)$}}
\end{overpic}
\label{fig:ts_deglognfit}
} \\
\subfloat[Empirical degree distribution of $G_1$ and fitted DPLN(2.75, 0.95, 2.4, 0.8)]{
\begin{overpic}[scale=0.5]{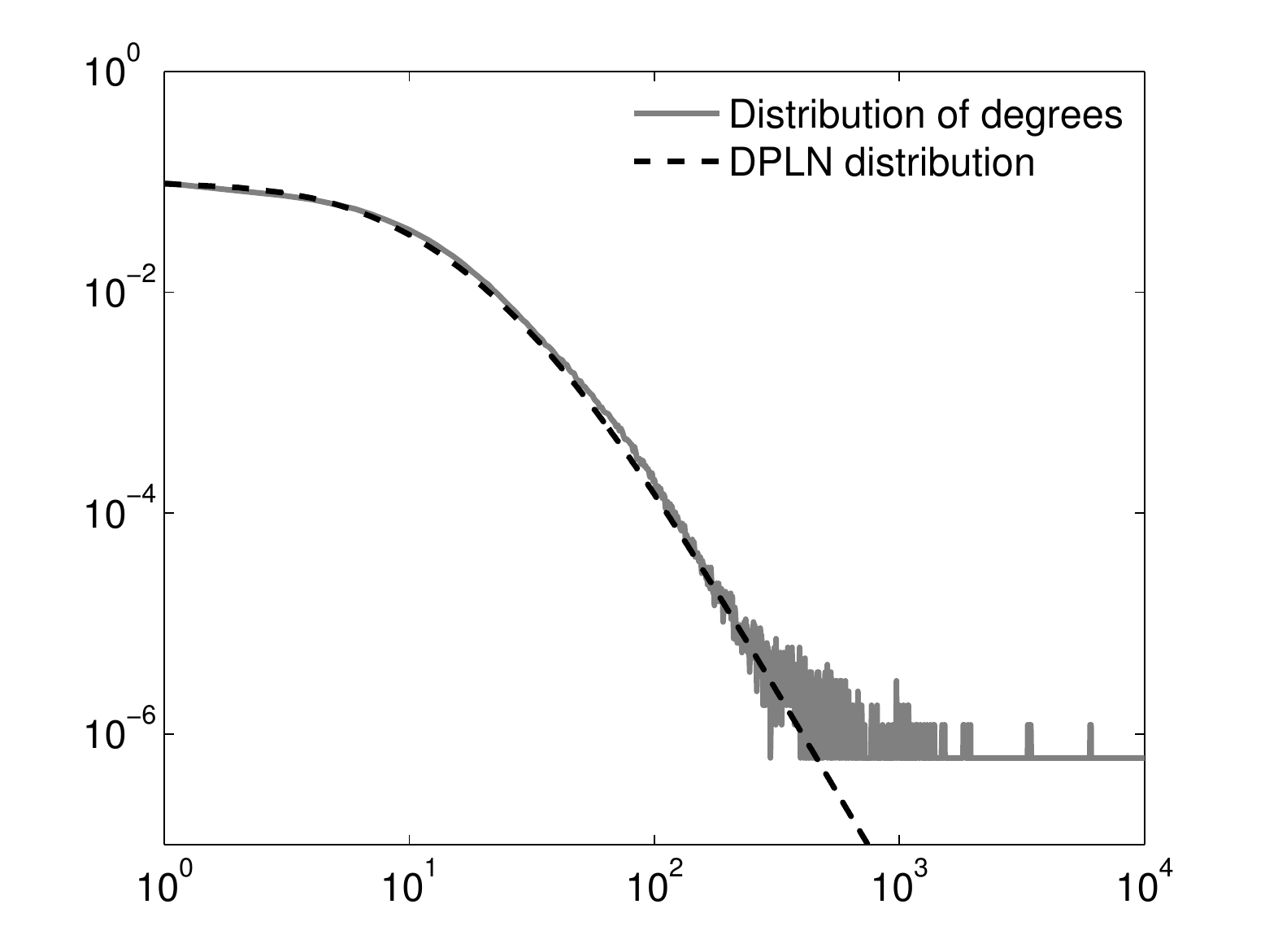}
 \put (43,0) {degree, $k$}
  \put (0,35) {\rotatebox{90}{$P(k)$}}
\end{overpic}
\label{fig:ts_pordegdplnfit}
} 
\subfloat[Empirical degree distribution of the reduced network $G_2$ and fitted LogN(2.15, 0.92)]{
\begin{overpic}[scale=0.5]{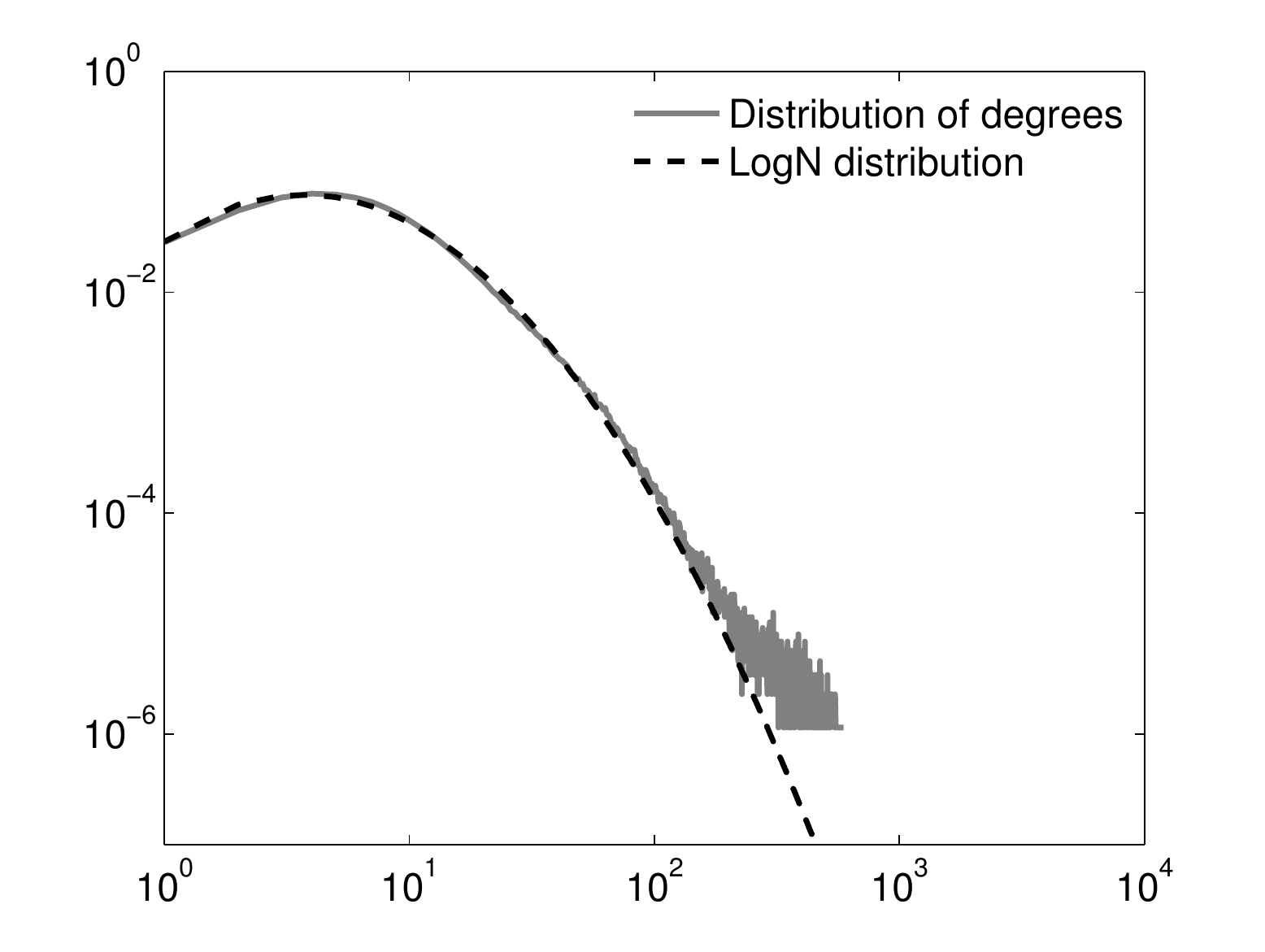}
 \put (43,0) {degree, $k$}
  \put (0,35) {\rotatebox{90}{$P(k)$}}
\end{overpic}
\label{fig:ts_pordeglognfit}
}
\endgroup
\caption{Degree distributions in the two studied networks, after 6 months of communications. Top: Belgium; bottom: Portugal}
\label{fig:ts_degdistfits}
\end{figure*}

Interestingly, we observe two different degree distributions in these networks: the degree distribution of $G_1$ seems to follow a Double Pareto LogNormal distribution (DPLN) while the sampled network $G_2$ shows a LogNormal degree distribution. Figure~\ref{fig:ts_degdistfits} shows the degree distributions of $G_1$ and $G_2$ and fitted DPLN and LogNormal for the Belgian and Portuguese datasets. \\

First, let us notice that the two datasets give qualitatively similar results. Distributions for both $G_1$ networks seem to correspond well to a DPLN distribution, while both distributions for $G_2$ networks present LogNormal behavior, indicating that this result may be universal across datasets. Furthermore, the parameters fitted for the distributions are of the same order of magnitude in both datasets. The small variations of parameters were to be expected since the sampling of the two datasets do not cover exactly the same proportion of the population, and since the two datasets come from different countries and have been recorded with different methods. Therefore, their characteristics may differ sufficiently to induce small discrepancies in the parameters. \\  

It is hardly surprising that the degree distributions of the two networks $G_1$ and $G_2$ are different. However, it is interesting to notice that there is a very close relationship between those two distributions, and that this relationship seems to be universal across different datasets. This relationship could explain some of the discrepancies between the observations published previously on networks of mobile phone datasets \citep{aiello2000random, seshadri2008mobile}. We now give a short description of the DPLN and of a process to produce these distributions, and for more details we refer the interested reader to the full paper on DPLN distributions \citep{reed2004double}.
The probability density function of the DPLN$(\alpha, \beta, \nu, \tau^2)$ distribution is of the form: 
\begin{equation}
\begin{aligned}
f(x) = \frac{\alpha \beta}{\alpha + \beta} & \left[  A(\alpha, \nu, \tau) x^{-\alpha-1} \Phi \left( \frac{\log x - \nu -\alpha \tau^2}{\tau}\right) \right. \\
&+ \left. x^{\beta-1} A(-\beta, \nu, \tau) \Phi^c \left( \frac{\log x - \nu + \beta \tau^2}{\tau} \right)\right], 
\end{aligned}
\end{equation}
with 
\begin{equation}
A(\theta, \nu, \tau) = \exp \left( \theta \nu + \theta^2 \tau^2/2 \right), 
\end{equation}
and where $\Phi$ and $\Phi^c$ represent the cumulative distribution function of the standard normal distribution, and its complementary distribution function, respectively. 
One way of generating a DPLN distribution is to observe a geometric brownian motion process with an initial condition drawn from a LogNormal distribution, but to stop the observation of this process at a random time. Indeed, if we suppose a variable $X$ evolves with a geometric brownian motion :
\begin{equation}
dX = \mu X dt + \sigma X dw,
\end{equation} 
where $t$ is the time variable and $w$ is a brownian motion, and if we suppose that $X$ has an initial condition drawn from a LogNormal distribution: $ X(0) \sim \textrm{logN}(\nu,\tau^2) $, then for any constant time $T$, $X(T)$ is distributed as a LogNormal with parameters depending on $T$ as: 
\begin{equation}
X(T) \sim \textrm{logN}(\nu + (\mu - \sigma^2/2)T, \tau^2+\sigma^2T).
\end{equation}
If, on the other hand, $T$ is drawn from an exponential distribution with probability density function  
\begin{equation}
f_{T} (t)  = \lambda e^{-\lambda t}, \ \ t >0, 
\end{equation}
then $X(T)$ exhibits a DPLN distribution \citep{reed2004double}.\\

In our case, to apply this situation to a mobile phone network, we let the variable $X$ correspond to the degree of a user, growing with time as the user makes new contacts. Further, we let variable $T$ correspond to the time between the first and last observations of activity of the user, representing the time during which the degree of the user could grow and be observed in our dataset. The degree distribution we observe in $G_1$ then corresponds to the distribution of $X(T)$. When $T$ is fixed and we observe users over a time window of fixed length, then we observe a LogNormal.\\

\section{Conjecture on the emergence of Double Pareto distributions}
We surmise that this process may explain the discrepancy between the degree distributions of networks $G_1$ and $G_2$: the LogNormal distribution corresponds to a network where no user entered or left the database during the period under study, so that users in the system all have the same age and are ate the same stage in the degree growth process. This observation is in line with other studies on different kinds of datasets such as, for example, the number of citations \citep{radicchi2008universality} or the number of votes in proportional elections \citep{fortunato2007scaling}. However, if we also consider users for which we have only partial information, that is, users that have left or entered the network during the observation period, then we observe a degree distribution corresponding to a DPLN, emerging as a mixture of many users of different age in the system, thus a mixture of many LogNormal processes, taken at different stages. \\

\begin{figure*}[t!]
\subfloat[]{\begin{overpic}[width=0.47\textwidth]{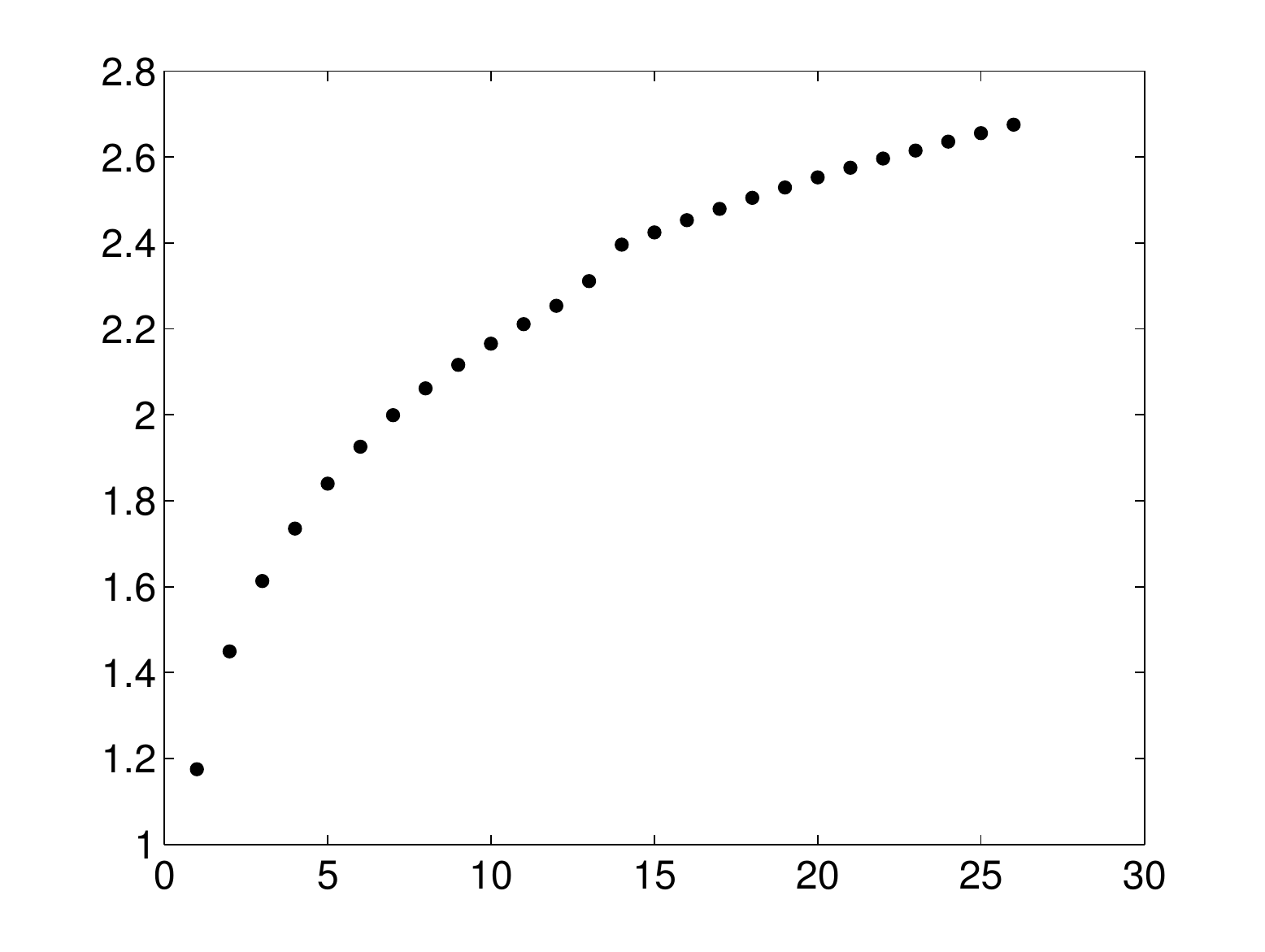}
 \put (18,-1) {length of the time window, in weeks}
  \put (0,35) {\rotatebox{90}{$\hat{\mu}$}}
\end{overpic}}
\subfloat[]{\begin{overpic}[width=0.47\textwidth]{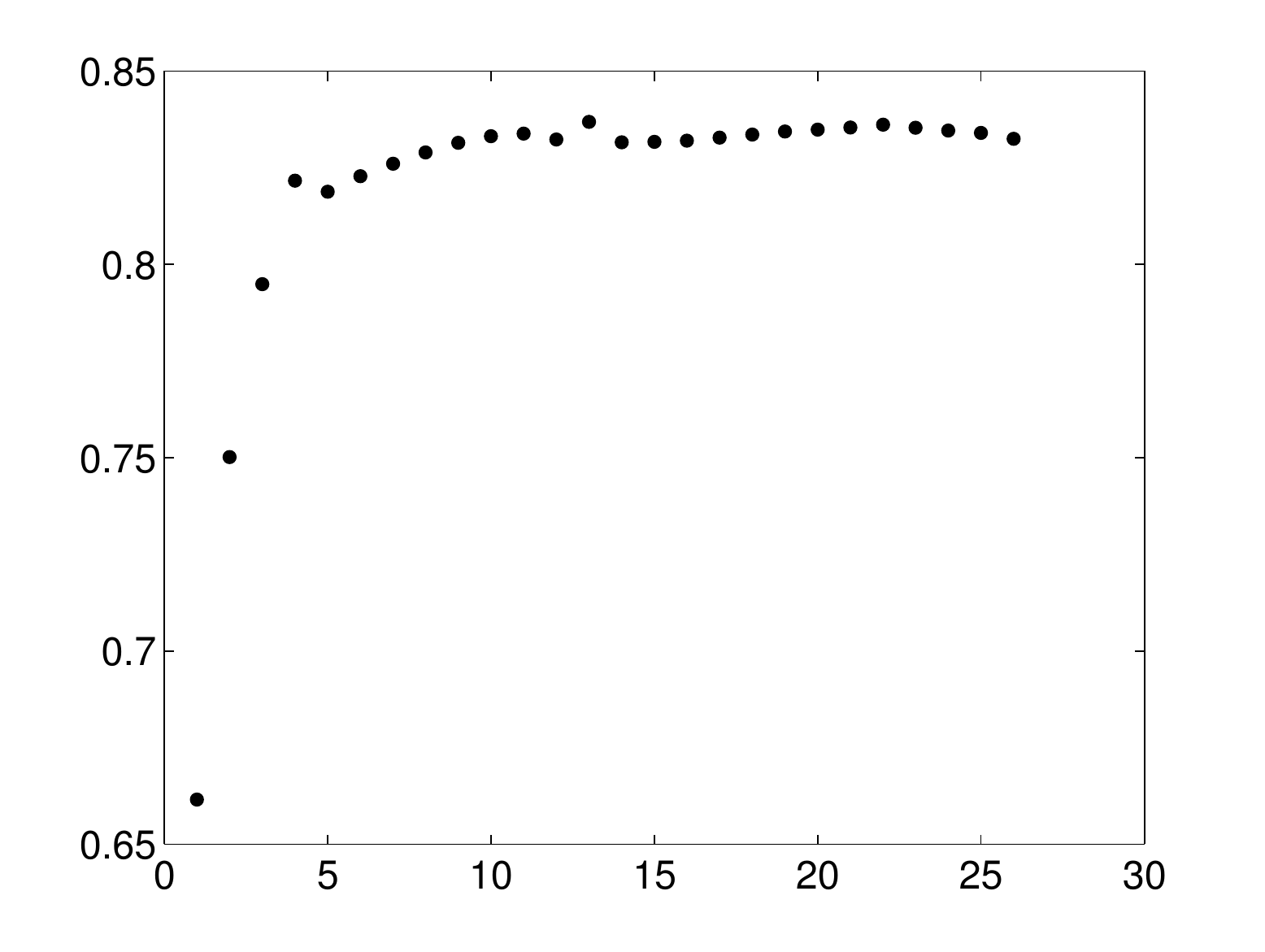}
 \put (18,-1) {length of the time window, in weeks}
  \put (0,35) {\rotatebox{90}{$\hat{\sigma}^2$}}
\end{overpic}}\\
\subfloat[]{\begin{overpic}[width=0.47\textwidth]{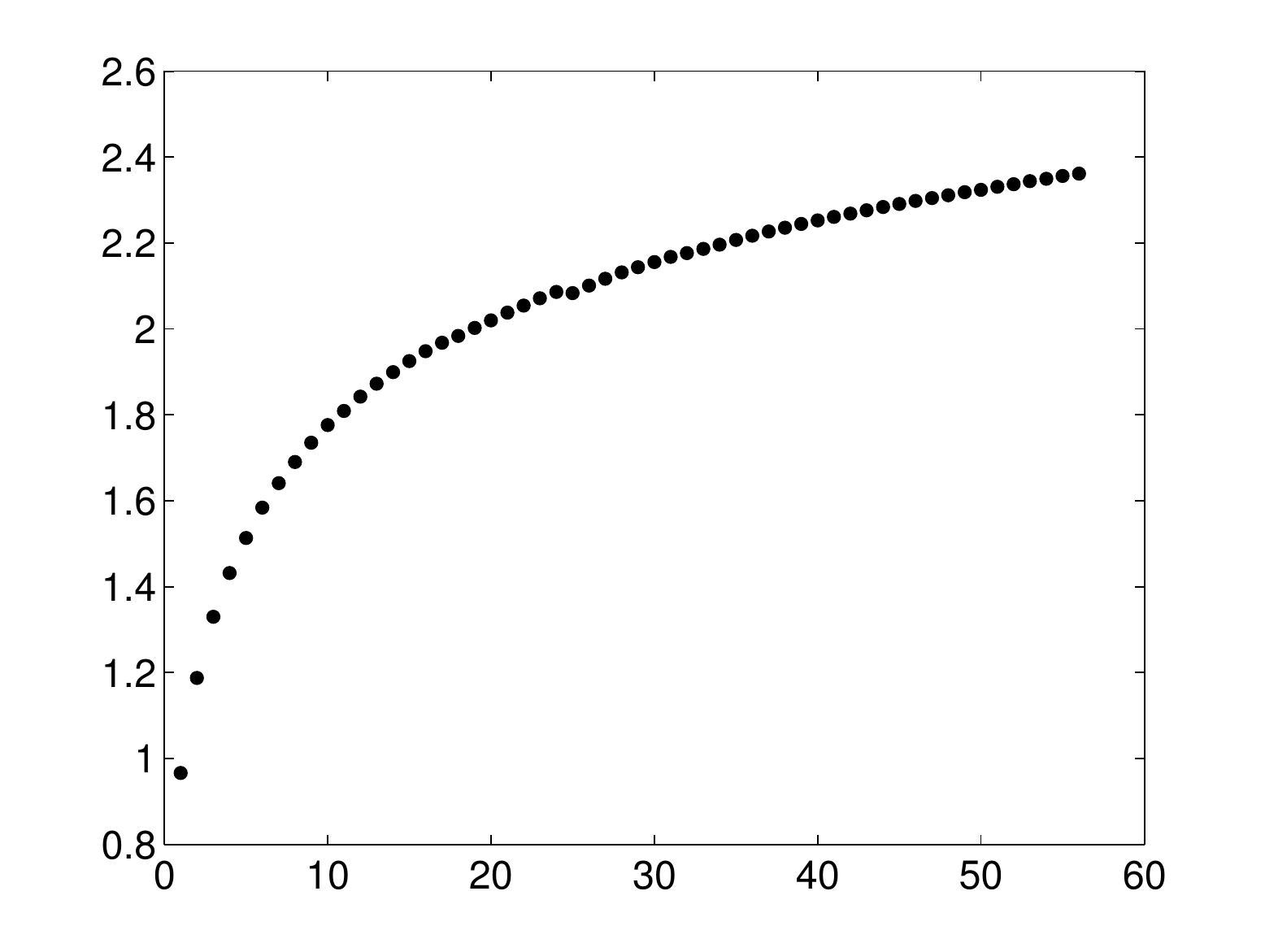}
 \put (18,-1) {length of the time window, in weeks}
  \put (0,35) {\rotatebox{90}{$\hat{\mu}$}}
\end{overpic}}
\subfloat[]{\begin{overpic}[width=0.47\textwidth]{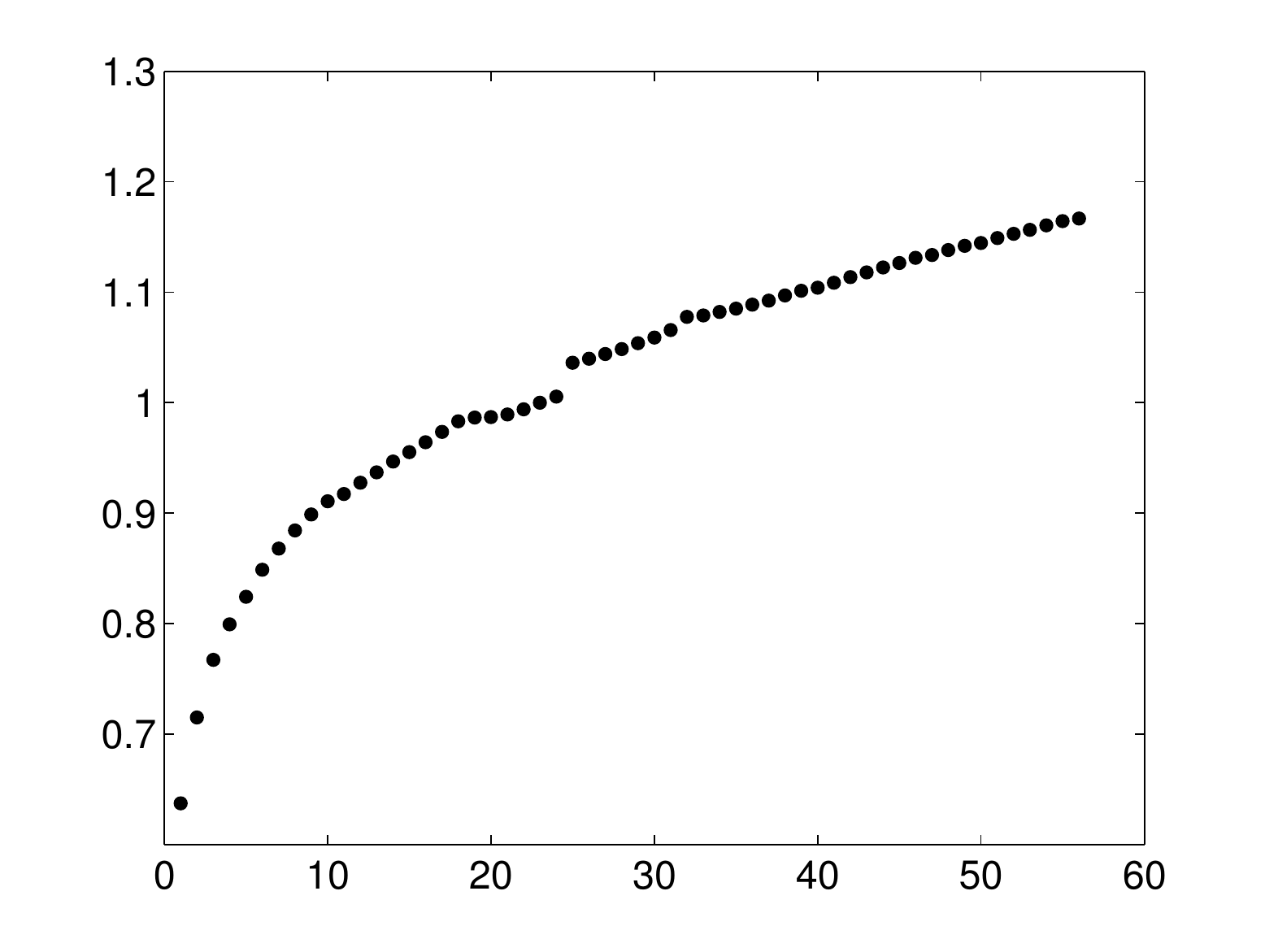}
 \put (18,-1) {length of the time window, in weeks}
  \put (0,35) {\rotatebox{90}{$\hat{\sigma}^2$}}
\end{overpic}}
\caption{Maximum likelihood parameters for the LogNormal distribution corresponding to the distribution of degrees of the graph $G_2$, depending on the length of the time window of observation (in weeks). Top: based on data for Belgium, bottom: data for Portugal}
\label{fig:ts_musig}
\end{figure*}
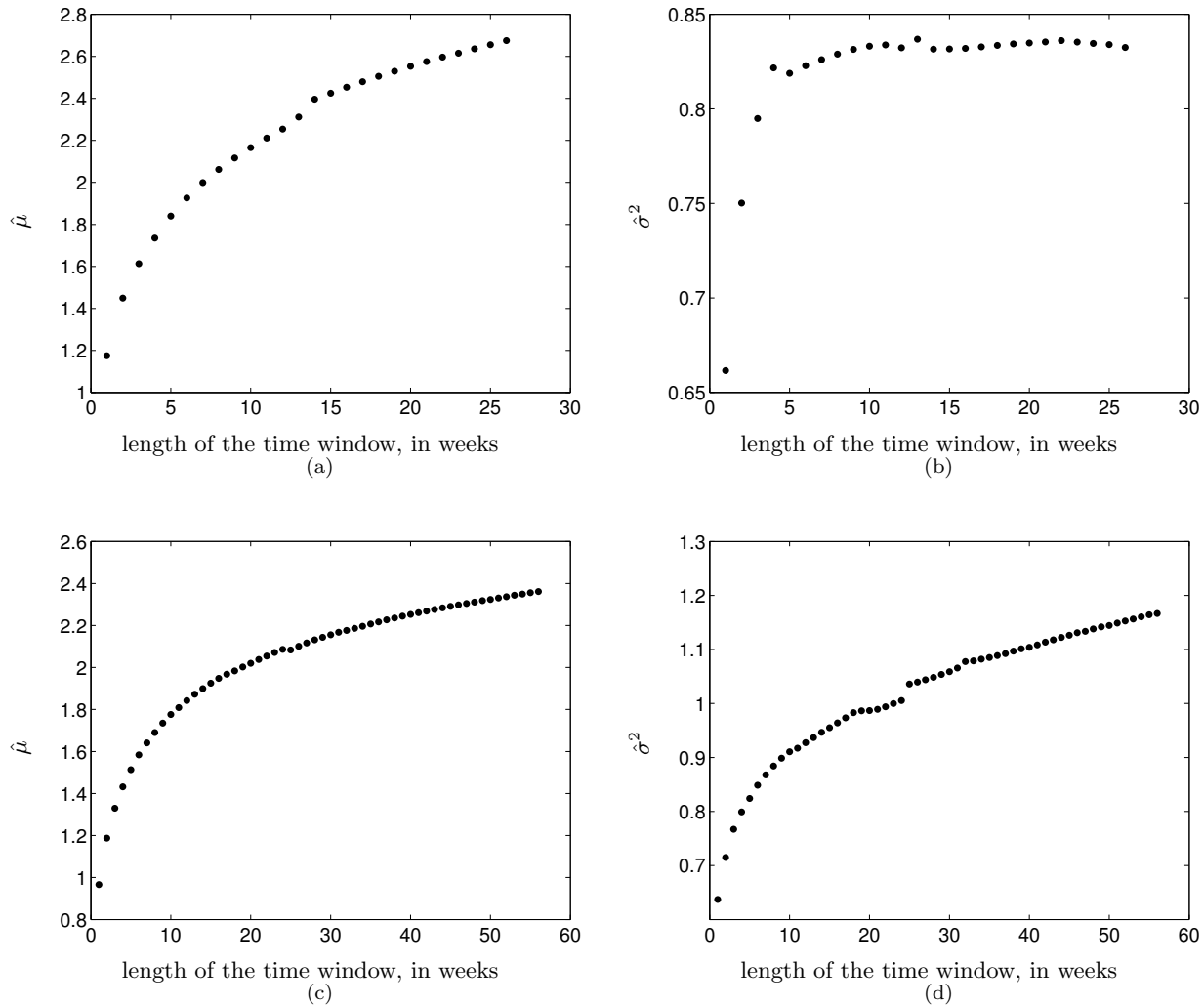

Empirically, when we look at a fixed number of users during a fixed length of time, the degree distribution obtained is a LogNormal (recall Figures \ref{fig:ts_deglognfit} and \ref{fig:ts_pordeglognfit}). 
Furthermore, the LogNormal distribution that fits the degree distribution evolves with the length of the time window of observation, which suggests that the process observed may correspond to the evolving process generating a DPLN. Figure~\ref{fig:ts_musig} shows the maximum likelihood parameters for the LogNormal distribution fitting the degree distribution of the network, for time windows of observation ranging from one week to 26 weeks for Belgium, and to 57 weeks for Portugal. After a transient behavior due to the start of the observation period, we observe that the parameters continue evolving with time. The Portuguese dataset covering a longer period, the observations are consistent with the hypothesis that the degree distribution is well approximated by a LogNormal whose parameters evolve linearly with time, as is described in the process generating DPLN distributions. In \citep{reed2004double}, this evolution is the consequence of a geometric brownian motion. In our analyses, we cannot validate or reject the hypothesis that the process corresponds to a geometric brownian motion, but instead we observe directly the evolution of the parameters of the LogNormal.  \\

Now if instead of looking at a fixed number of people, we let people enter and leave the network during the observation window, this is equivalent to observing the whole population, but during a time-window that is different for each individual. Indeed, the time when a user switches to or from another network determines the time frame during which this user was observed. Furthermore, when a new user enters the network, they start with a degree equal to zero, whereas the degree of a user is frozen as soon as they leave the network and are no longer observed. Therefore, if we assume that every user has a degree evolving according to a LogNormal distribution with parameters evolving with time (as empirically validated above), then the degree distribution that we observe when taking all users into account corresponds to a mix of LogNormal distributions with various parameters, yielding the distribution that we observe for $G_1$, which is well fitted by a DPLN. 
Let us note that for a short time window during which users have not yet had the time to enter or leave the network, the distribution for $G_1$ still corresponds to a LogNormal distribution.\\ 

\begin{figure}[hbt]
\centering
\subfloat[]{
\begin{overpic}[scale=0.45]{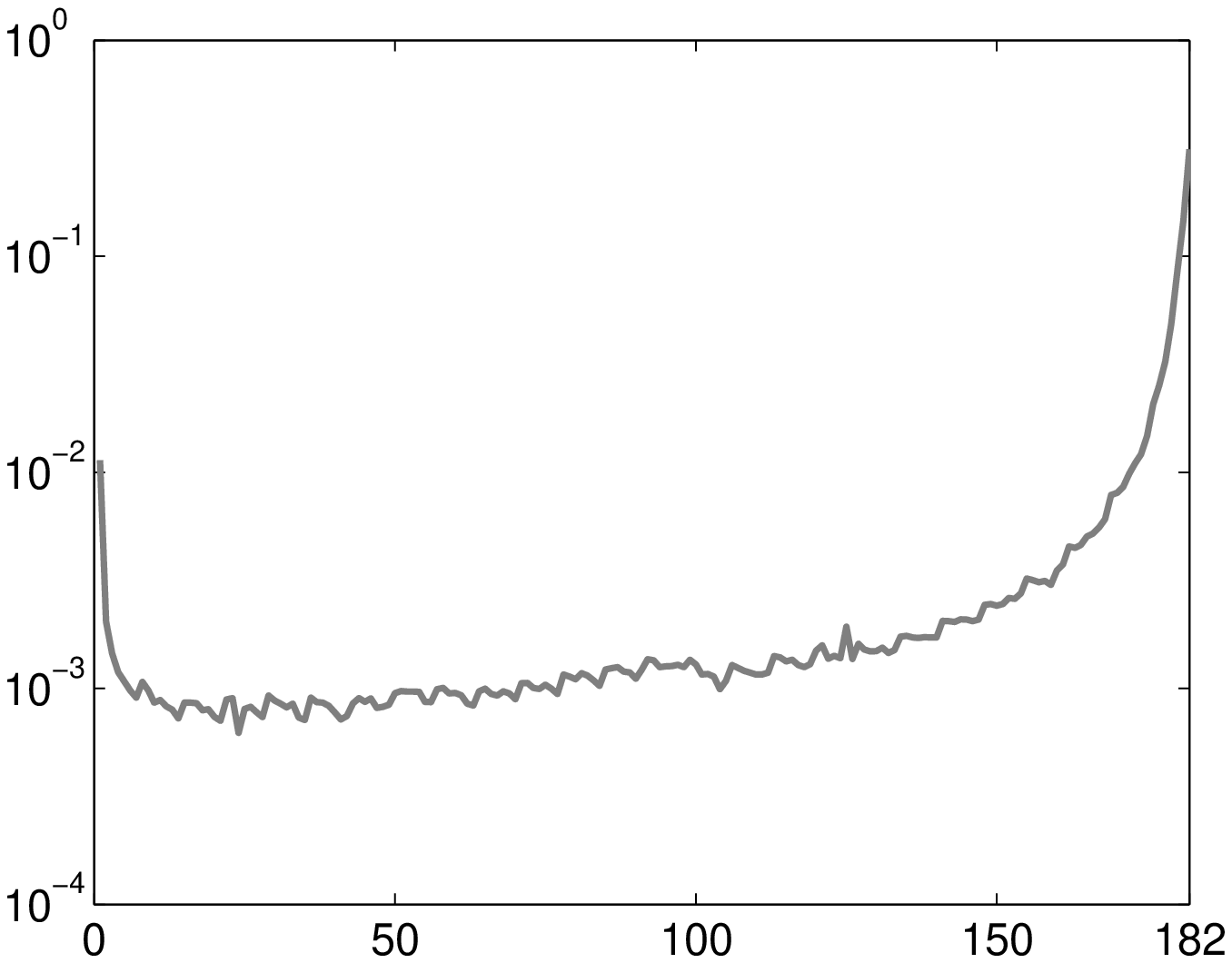}
 \put (47,-1) {time, $t$}
  \put (0,35) {\rotatebox{90}{$P(t)$}}
\end{overpic} }\\
\subfloat{
\begin{overpic}[scale=0.45]{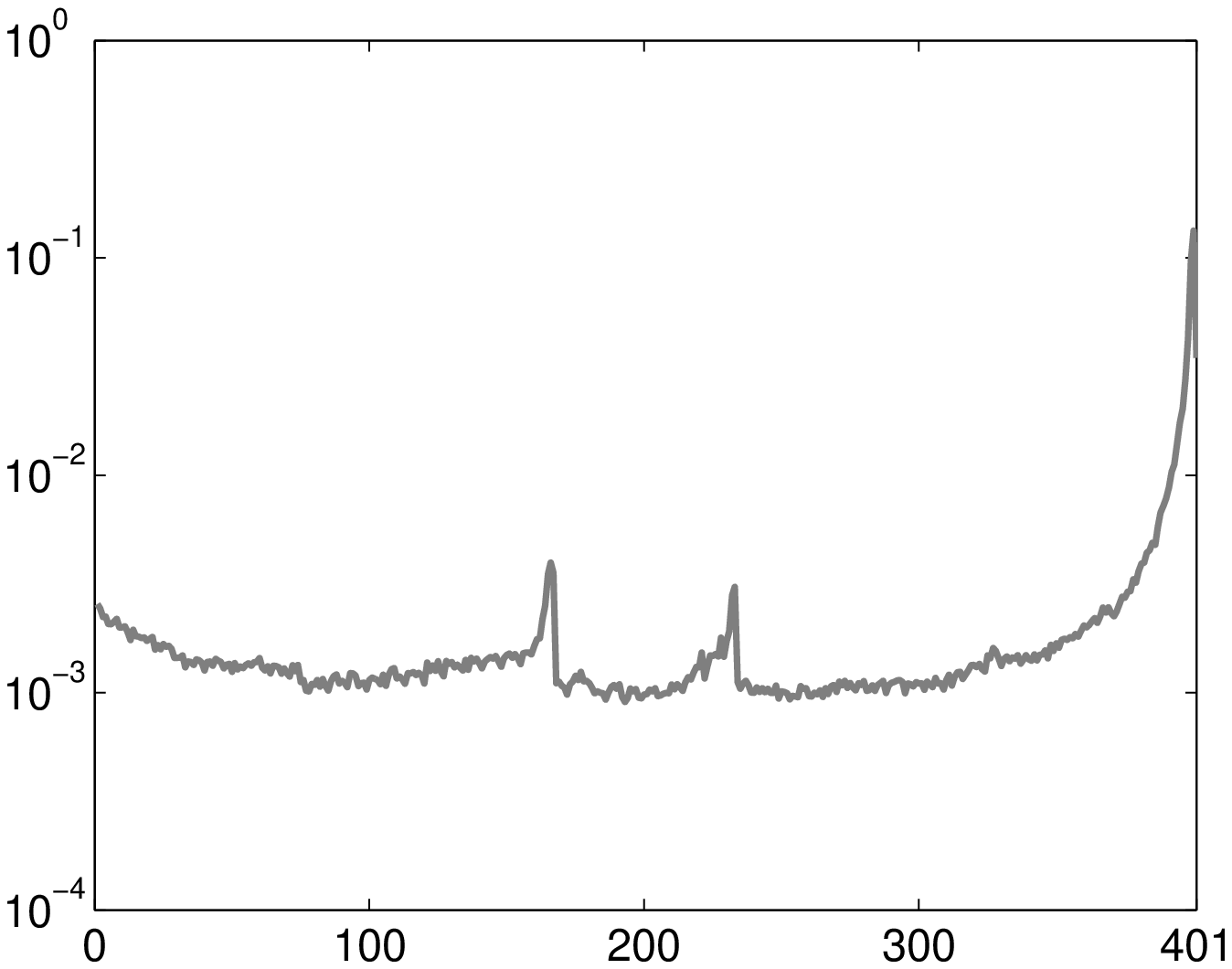}
 \put (47,-1) {time, $t$}
  \put (0,35) {\rotatebox{90}{$P(t)$}}
\end{overpic} }
\caption{Distribution of elapsed times between first and last observed activity in the network (in days). Top: Belgium, bottom: Portugal. The two peaks for the Portuguese dataset after five and a half months, and after eight months are due to the missing 45 days of data: users that have left the network during this period all appear to have left the network on the same last day before the gap, and similarly for the users who joined the network during the gap period, they all appear to have joined on the first day after the gap.}
\label{fig:ts_timeinthenet}
\end{figure}

However, the generating process described in \citep{reed2004double} supposes that the distribution of times during which users have been observed corresponds to an exponential distribution. Unfortunately, this is not the case in the networks under study here. The distribution of elapsed time (in days) between the first and last activity is depicted on Figure \ref{fig:ts_timeinthenet}, for Belgium (top) and Portugal (bottom). We observe that these distributions are strongly influenced by the maximum time determined by the size of the dataset (182 and 401 days of data, respectively). Furthermore, even if we only take into account the left part of the distribution in the Belgian dataset (between 0 and 50 days), the distribution is still too broad to correspond to an exponential curve. In the case of Portugal, the distribution of observation times shows even a less pronounced peak for small times. \\

These results suggest that the generating process of a DPLN distribution may be more robust than suggested by the process uncovered in \citep{reed2004double}. We therefore suggest a new conjecture: the generating process of the DPLN distribution admits the relaxation of the hypothesis that the time of observation is distributed exponentially. \\

\begin{figure*}[t!]
\subfloat[distribution of $t$ drawn from a combination of three uniform distributions]{\begin{overpic}[width=0.47\textwidth]{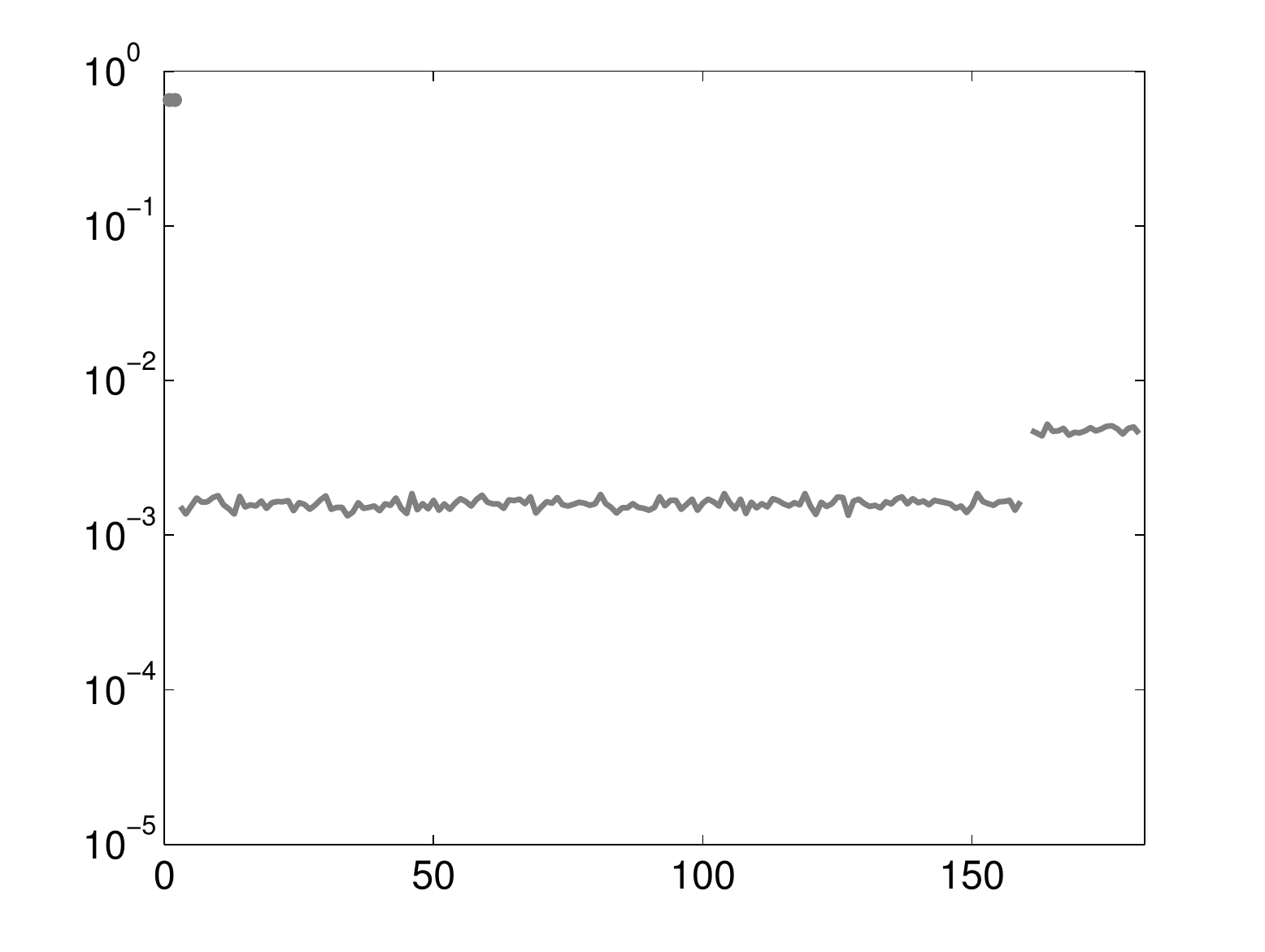}
 \put (43,-1) {time, $t$}
  \put (0,35) {\rotatebox{90}{$P(t)$}}
\end{overpic}}
\subfloat[synthetic degree distribution]{\begin{overpic}[width=0.47\textwidth]{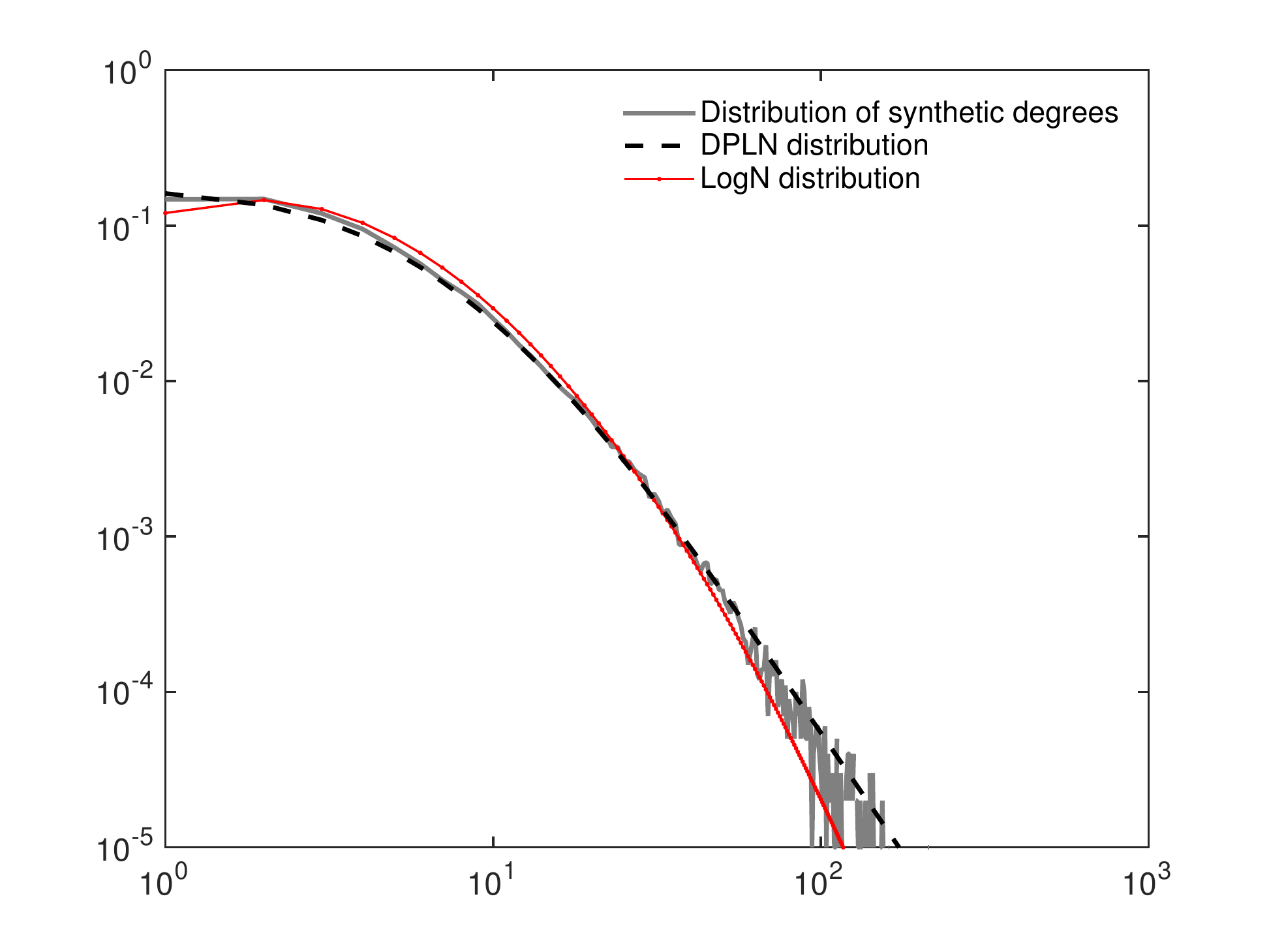}
 \put (43,-1) {degree, $k$}
  \put (0,35) {\rotatebox{90}{$P(k)$}}
\end{overpic}}\\
\subfloat[distribution of $t$ drawn from a sum of exponential distributions]{\begin{overpic}[width=0.47\textwidth]{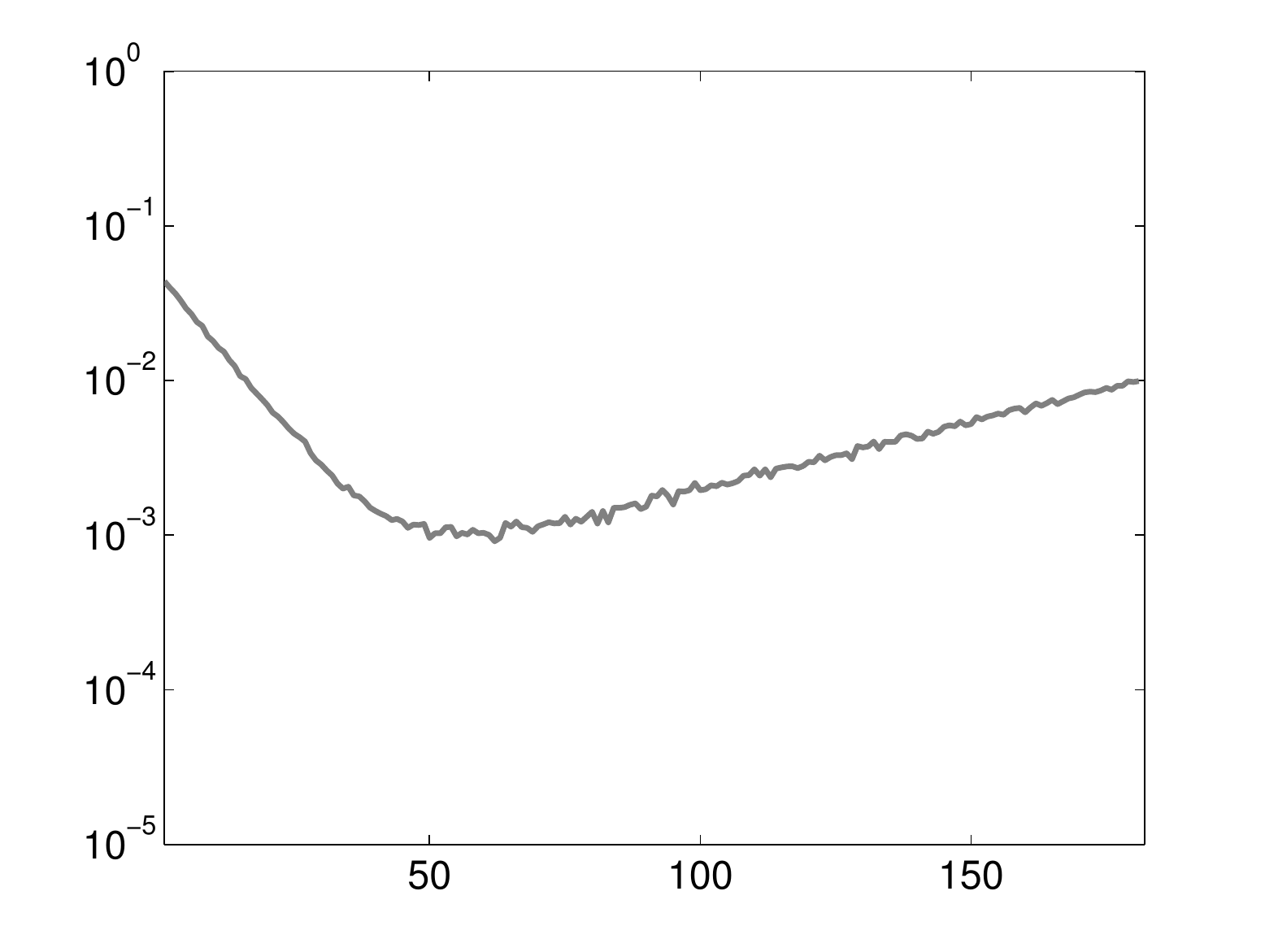}
 \put (43,-1) {time, $t$}
  \put (0,35) {\rotatebox{90}{$P(t)$}}
\end{overpic}}
\subfloat[synthetic degree distribution]{\begin{overpic}[width=0.47\textwidth]{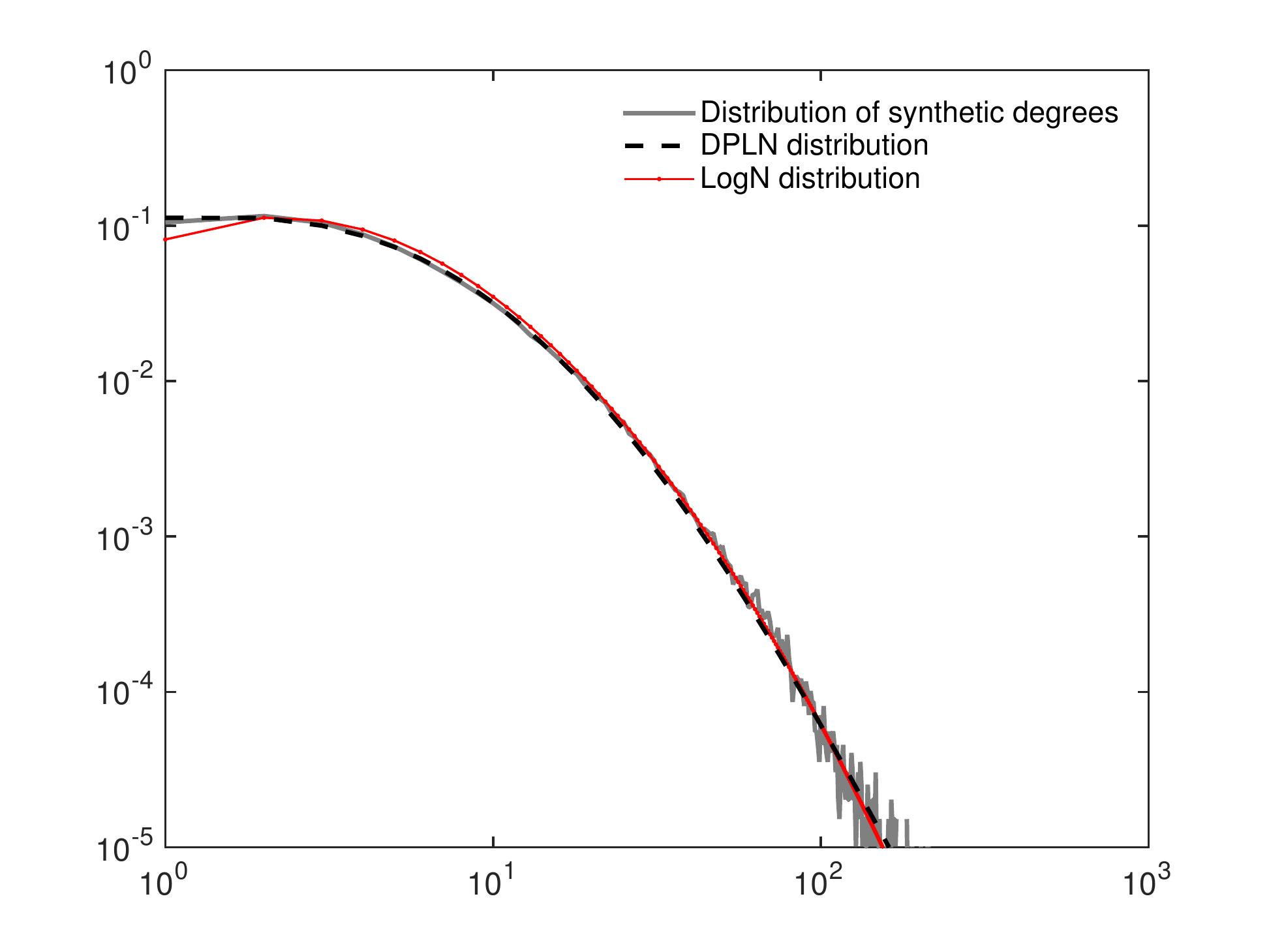}
 \put (43,-1) {degree, $k$}
  \put (0,35) {\rotatebox{90}{$P(k)$}}
\end{overpic}}
\caption{On the left: distributions chosen for $T$. (a) combination of three uniform distributions : high probability for very short times (1 to 3 days), low probability for medium times (up to 160 days), and a high probability again for large times before the cutoff (180 days). (c) sum of two exponential distributions: one growing exponential with cutoff after 180 days, and one decaying exponential distribution. On the right: synthetic degree distributions generated by reproducing the process of generation of DPLN distributions but with variable $T$ drawn from the distributions shown on the right, along with the curves of fitted DPLN and LogNormal distributions.}
\label{fig:ts_variousdpln}
\end{figure*}

To strengthen our hypothesis, we conduct several Monte-Carlo experiments, reproducing the process described above, with different distributions for choosing $T$. For each experiment, we randomly select 100,000 values of $T$ from a chosen distribution (different than exponential), and then select a value for the degree $k_i$ from a LogNormal distribution whose parameters depend on the value selected for $T_i$ and grow linearly with $T_i$. We then observe the resulting distribution of synthetic degrees $k$. Figure~\ref{fig:ts_variousdpln} shows the results of our simulations for two different distributions of the variable $T$: a combination of uniform distributions, and a sum of exponential distributions. These distributions were chosen as they reproduce some of the features observed in the empirical distributions observed on mobile phone data: a peak for very short times and a peak at the maximum time (as limited by the observation time window, see Figure~\ref{fig:ts_timeinthenet}). We observe that the distributions obtained for $k$ are better approximated by DPLN distributions than by LogNormal distributions, especially for small values of $k$. Furthermore, a test for normality of the logarithm of the samples rejected the null hypothesis that those samples correspond to LogNormal distributions (associated p-values $<10^{-3}$).
These results suggest that DPLN distributions may arise as a result of mixing LogNormal distributions with time-evolving parameters, but the hypothesis that the time of observation must be distributed exponentially may not be necessary for a DPLN distribution to appear. 

\section{Discussion and conclusion}

We have presented an observation from the analysis of two large databases of mobile phone communications in Belgium and Portugal. We showed that the measurement of the degree distribution of a time-evolving social network may be more complicated to analyze than initially thought, as we showed that seemingly benign choices of data cleaning such as the inherent sampling of the data can lead to radically different conclusions. 
We also showed that the Double Pareto LogNormal distribution, observed in several studies as a characteristic of social networks does not represent the distribution of the underlying social network, but instead appears as an effect of the sampling of users and of the limited time window, the actual degree distribution corresponding rather to a LogNormal. Moreover, we showed that this effect appears in two different datasets indicating suggesting the tempting hypothesis calling for further verification that this is a universal characteristic of mobile phone datasets. Furthermore, we presented results indicating that the Double Pareto LogNormal distribution can arise with a process that does not exactly correspond to those studied before, opening the question of the robustness of this process for further study.  \\

\begin{figure*}[t!]
\centering
\subfloat[network $G_1$, Belgium]{\includegraphics[width=0.47\textwidth]{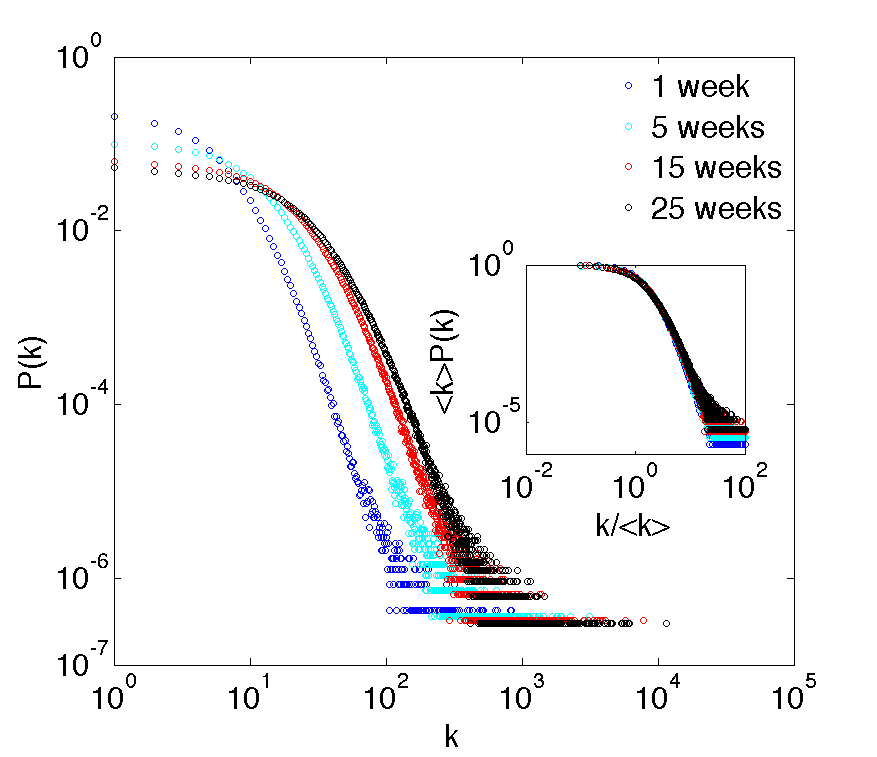}} 
\subfloat[network $G_2$, Belgium]{\includegraphics[width=0.47\textwidth]{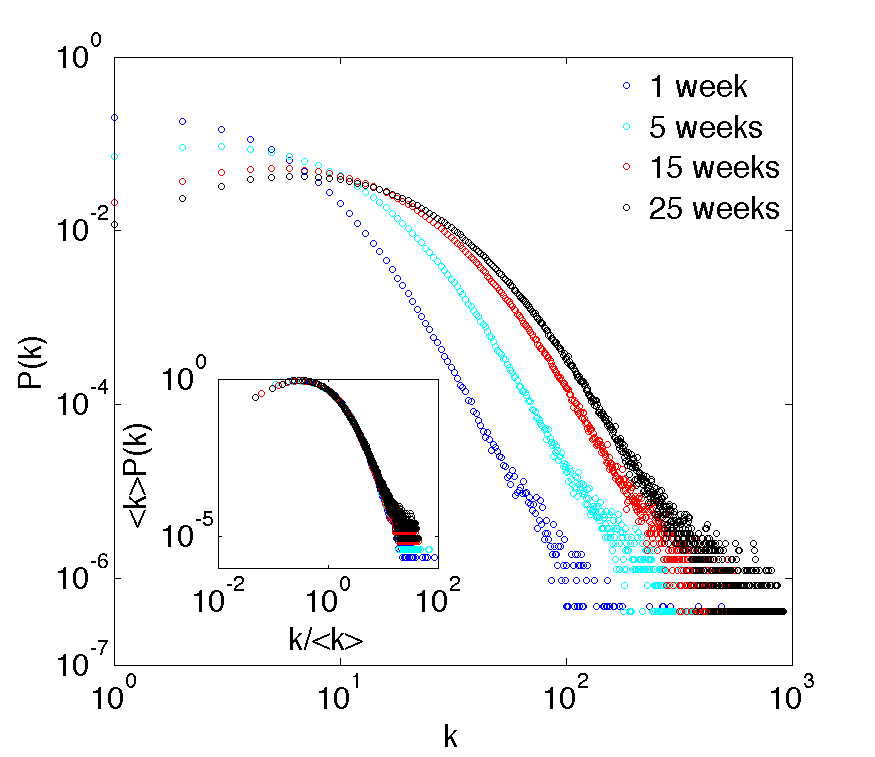}}
\\
\subfloat[network $G_1$, Portugal]{\includegraphics[width=0.47\textwidth]{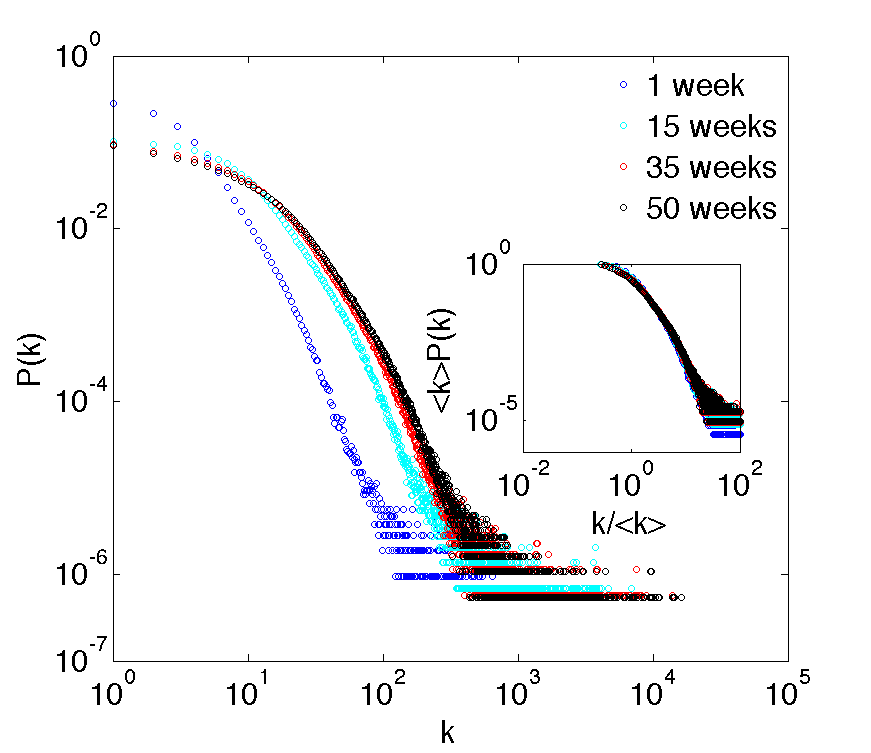}} 
\subfloat[network $G_2$, Portugal]{\includegraphics[width=0.47\textwidth]{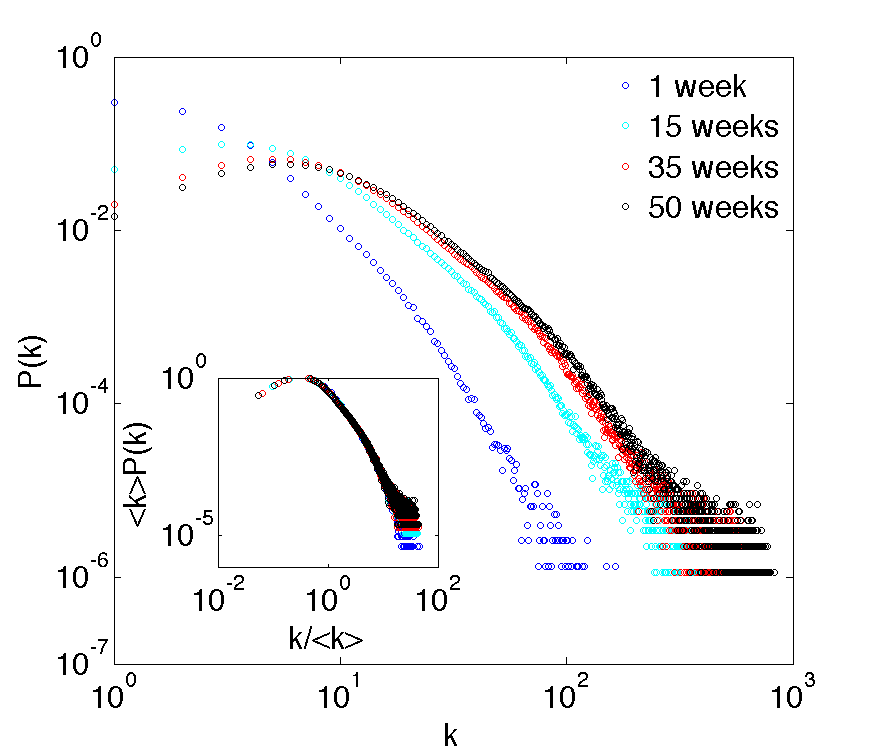}}
\caption{Degree distributions for different lengths of the observation time window of the network (a,c) on the network of all users $G_1$, and (b,d) on the network $G_2$ of users that stay active throughout the whole observation period. Top: Belgium, bottom: Portugal.}
\label{fig:ts_degdists}
\end{figure*}

Additionally, we observe that the degree distributions present similar characteristics whether we look a one or many weeks of data. Up to a scaling parameter all curves correspond to the same distribution, see Figure \ref{fig:ts_degdists}. Interestingly, we observe that the average degree is an appropriate estimator of this scaling parameter. This scaling property had already been observed in \citep{krings2012effects} on the network of all users $G_1$ of Belgium. Here, we also extend this observation to another dataset, and to the network of active users $G_2$ presenting a LogNormal degree distribution. \\

These observations suggest that the degree distribution of a mobile phone network may be governed by a universal law, and can be represented by a LogNormal distribution in the case of a constant population being observed, and by a DPLN in the case where an evolving population is being observed. Independently of the dataset, or of the length of the time window (provided that it is longer than one week), we observe that the degree distributions remain qualitatively the same, up to a scaling parameter. However, this may not be valid for very short time windows of the order of days, as was observed by Krings \emph{et al.}\ in \citep{krings2012effects}.
Nevertheless, this degree distribution law appears to be fairly robust to the particularities of specific datasets, as we have reached the same conclusions studying two different datasets, which do not cover the same population nor the same country. 

These results provide new insights into the particular characteristics of mobile phone datasets, and the dynamics of the construction of a social network based on mobile communications. To overcome some of the privacy issues related to the use of mobile phone datasets, recent work has focused on creating synthetic data reproducing the characteristics of empirical data \citep{isaacman2012human, mir2013dpwhere}. The observations presented in this paper may help in the future to create better synthetic datasets offering a closer correspondence with empirical data, as these results reveal characteristics of mobile phone datasets that may have been overlooked in the past. \\

The present study opens the door to many questions to be investigated regarding the effects of sampling of a dataset. Furthermore, the sources of bias in a dataset are numerous. We have discussed the sampling of the dataset, which is determined by which operator provided the data, and which part of the population has chosen that operator. If the operator is more popular across certain groups, determined by for example age, revenue or occupation, the coverage of the dataset may be biased. Moreover, this bias is very difficult to remove without access to a dataset with a perfect coverage, which does not exist. The question of the impact of sampling on the results of the analyses remains an open question, that would require further work. 
Further than the choice of provider and its associated sampling, additional sources of bias in the data include the behavior of users. For example, flashing techniques consist in letting a relative's phone ring a couple of times and waiting for them to call back. Such technique allows to insure that it is always the same person who pays for a communication, but links applying this technique will be removed from the dataset if filtering techniques impose reciprocity. Another example is given by people who prefer voice calls to text messages, or the opposite. If only one type of communication is recorded in a dataset, some links appear imbalanced because the two nodes have different preferences. In addition, let us note that a single sim card may be shared between several people while other users possess more than one phone number, and that detecting these users is a difficult task. 
Depending on the specific filtering methods used prior to the analysis of the data, such type of behavior of the users may induce additional bias in the extracted social network. 
However, so far it is not clear how to evaluate the sources and the extent of biases that are present in a given dataset, and researchers must pay attention to the interpretation of their results, bearing in mind that the data analyzed are far from being perfect.

\section{Acknowledgements}
AD acknowledges funding from Fonds National de la Recherche Scientifique (F.R.S. - FNRS). This research was made possible with the support of Orange. We acknowledge support from a grant ``Actions de
recherche concert\'ees -- Mining and Optimization of Big Data Models" of the ``Communaut\'e Fran\c caise
de Belgique" and from the Belgian Network DYSCO (Dynamical Systems, Control,
and Optimization), funded by the Interuniversity Attraction Poles Programme, initiated
by the Belgian State, Science Policy Office. We also acknowledge support from Innoviris in the context of the Bru-Net project.


\end{document}